\documentclass[%
reprint,
amsmath,amssymb,aps,
]{revtex4-1}

\usepackage{graphicx}% Include figure files
\usepackage{dcolumn}% Align table columns on decimal point
\usepackage{bm}% bold math

\begin{document}

\preprint{APS/123-QED}

\title{Impact of density-dependent migration flows \\ on epidemic outbreaks in heterogeneous metapopulations}

\author{J. Ripoll}
\author{A. Aviny\'o}
\author{M. Pellicer}%
\author{J. Salda\~{n}a}
\affiliation{%
 Departament d'Inform\`atica, Matem\`atica Aplicada i Estad\'{\i}stica, Universitat de Girona, 17071 Girona, Catalunya (Spain)
}%

\date{\today}% It is always \today, today,

\begin{abstract}
We investigate the role of migration patterns on the spread of epidemics in complex networks. We enhance the SIS-diffusion model on metapopulations to a nonlinear diffusion. Specifically, individuals move randomly over the network but at a rate depending on the population of the departure patch. In the absence of epidemics, the migration-driven equilibrium is described by quantifying the total number of individuals living in heavily/lightly populated areas. Our analytical approach reveals that strengthening the migration from populous areas contains the infection at the early stage of the epidemic. Moreover, depending on the exponent of the nonlinear diffusion rate, epidemic outbreaks do not always occur in the most populated areas as one might expect. \medskip
\end{abstract}

%\pacs{Valid PACS appear here ...}% PACS, the Physics and Astronomy
\maketitle

\section{Introduction}

Density dependence is a crucial factor in dispersal movements of individuals in many species, be they animal or human. In turn, these movements are a key factor in the fate of epidemic outbreaks occurring within spatially distributed populations \cite{Meloni}.

Migration or dispersal, as the movement of individuals from one place to another, are affected by the presence of other conspecific members in two different ways. A positive density dependence reflects the fact that high population densities increase competition effects among their individuals and induce emigration from heavily populated locations. On the contrary, a negative density dependence corresponds to conspecific attraction, i.e., the tendency for individuals of a species to settle near conspecifics, and it has been claimed to be one of the reasons for leaving low-density areas \cite{Matthysen} and for population aggregation patterns \cite{Mendez}. In general, there is empirical evidence that diffusion rates can both increase and decrease with population density. For instance, in some insect groups, crowded conditions lead to the appearance of a greater fraction of long-winged adults \cite{Kisimoto} and, in aphids, the appearance of winged individuals within populations of wingless adults \cite{Harrison}. On the other hand, negative density dependence has been reported, for instance, in gulls, voles, and deers \cite{Matthysen}. In humans, both positive and negative dependences have been considered for prehistoric and historical human population dispersals \cite{Steele2009}.

Traditionally, populations are assumed to be continuously distributed over space and, hence, nonlinear reaction-diffusion equations have been used to study the effect of the previous density dependencies on the spatial population dynamics \cite{Murray, Mendez}. In this context, linear dependence of the diffusion coefficient on the population density has been proposed by several authors for animal dispersal \cite{Gurney1975, Shigesada}, whereas a power law with positive exponent has been used to describe the relationship between the insect dispersal rate and the population density \cite{Murray}.

Alternatively, the increasing fragmentation of habitats of many species, as well as the fact that, at a large scale, human travel can be described by flows among a set of discrete locations, make metapopulation models very useful for studying the dynamics of spatially subdivided populations and, also, for the analysis of spreading processes on top of these populations. Under this modeling approach, the spatial distribution of the whole population is described by a network of patches inhabited by local populations, as cities or habitats in a patchy landscape, with migratory flows connecting them.

The nature of these flows is, in fact, a key ingredient of metapopulation  models. On the one hand, flows can be due to uniformly random migration, which assumes that any neighbouring patch of a ``source'' patch is reached with the same probability. In this case, the migration probability only depends on features of the source patch as, for instance, its population density or its degree or connectivity. On the other hand, migratory flows between patches can depend on features of both source and destination patches. Such a non-uniformly random migration is, in fact, assumed by the so-called gravity model of movements of people and goods traditionally used in social sciences \citep{Brockmann2009, Wilson}, and by the radiation model introduced in \cite{Simini2012} to overcome some of its limitations. In an epidemic context, migration towards other patches can be even more related to the healthy conditions in the destination patches than to the conditions in the origin patch. In \cite{Wang-2012}, for instance, a constant diffusion rate is assumed for each class of individuals, but the migration probability between two patches depends on the susceptible population in the destination patch.

The goal of this paper is to deal with positive and negative density-dependent diffusion processes extending the equations for continuous-time epidemic dynamics on metapopulations derived in \cite{Saldana2008}, which were based on the formalism of complex networks introduced in \citep{Colizza2007, Colizza-Vespignani-2007}. In particular, we study the impact of density-dependent diffusion coefficients on the population distribution among heavily populated and  lightly populated areas, how these distributions affect the epidemic growth, and the contribution of each local population to the epidemic spreading.

\section{The model}

The spatial arrangement of patches (areas) is described in terms of the connectivity (degree) distribution $p(k)$ and the conditional probability $P(k'|k)$ for a patch (node) of degree $k$ to be connected to a patch of degree $k'$. Moreover, within a patch of degree $k$, any individual has the same probability of leaving it through any of its $k$ links, namely, $1/k$, which implies that the strength of the connections is independent of the travel distance between patches.

According to these assumptions and denoting by $\rho_{S,k}(t)$ and $\rho_{I,t}(t)$ the average number of susceptible and infectious individuals in a patch of degree $k$ ($k=~1,~\dots,~k_{\max}$), respectively, the following equations describe the epidemic dynamics over a metapopulation
\begin{equation}\label{edo}
{\left\{
\begin{array}{l}
\rho'_{S,k}(t) \, = \, \left(\mu -
\beta c(\rho_k) \frac{\rho_{S,k}}{\rho_k}\right)\rho_{I,k} + \delta (\rho_k - \rho_{S,k}) \smallskip\\
\qquad \qquad - D_S( \rho_{k} ) \rho_{S,k}
{\displaystyle + k \sum_{k'}} D_S( \rho_{k'} ) P(k'|k) \frac{\rho_{S,k'}}{k'}\; ,
\medskip\\
\rho'_{I,k}(t) \, = \,  \left(\beta c(\rho_k) \frac{\rho_{S,k}}{\rho_k}- \mu \right)\rho_{I,k} - \delta \rho_{I,k}  \smallskip\\
\qquad \qquad - D_I ( \rho_{k} ) \rho_{I,k}
{\displaystyle + k \sum_{k'}} D_I( \rho_{k'} ) P(k'|k) \frac{\rho_{I,k'} }{k'} \; ,
\end{array}
\right.}
\end{equation}
where $\beta$ denotes the infection transmission probability through an infectious contact, $\mu$ is the recovery rate, and equal birth and death rates $\delta$ are assumed. The average population size in a patch of degree $k$ is $\rho_k(t)= \rho_{S,k}(t)+\rho_{I,k}(t)$, and the contact rate $c(\rho_k)$ is a non-decreasing density-dependent function, generalizing the two cases considered in \cite{Saldana2008, Juher2009}: $c(\rho) = \rho$ (fully-mixed population) and $c(\rho) = 1$ (limited homogeneous mixing). Finally, the density-dependent diffusion rates of susceptible and infectious individuals are denoted by $D_S( \rho_{k} )$ and $D_I( \rho_{k} )$, respectively. Note that it is natural to assume that the total outflow of individuals of each type in a patch, $D_S(\rho)\rho$ and $D_I(\rho)\rho$, are zero when $\rho = 0$ (see Discussion for a comment on the implications of the violation of this hypothesis).

The first term in each equation of (\ref{edo}) corresponds to the infection and recovery processes. The second term is the neutral demographic turnover. The last one corresponds to the migration/diffusion process which can be split into negative and positive terms. The former count the number of individuals leaving a patch of degree $k$, whereas the latter are the sum of the flows of individuals arriving at this patch from patches of degree $k'$, provided that patches of degree $k$ are connected to patches of degree $k'$, i.e., for those $k'$ such that $P(k'|k) > 0$.

Once we know the solution of the system (\ref{edo}), the (expected) total number of susceptible and infected individuals are $S(t) =  N\, \sum_k p(k)\, \rho_{S,k}(t)$ and $I(t) = N\, \sum_k p(k)\, \rho_{I,k}(t)$, respectively, where $N$ is the number of nodes of the network.
From Eq.~\eqref{edo} and assuming the consistency condition $k P(k'|k) p(k) = k' P(k|k') p(k')$, it follows that the total number of individuals is conserved in the metapopulation, i.e., $S(t) + I(t)= N \rho^0$, with $\rho^0$ being the average number of individuals per patch.

%\subsection{Density-dependent rates} \label{sec:rates}

%Moreover, we will assume that the total outflow of individuals in any patch, $D_i(\rho_k) \rho_{i,k}$ with $i=I,S$, is a strictly increasing function of its population $\rho$. This implies $\alpha_i > -1$.

%\subsection{Equations in matrix form} \label{sec:matrix}

Denoting by $\vec{\rho}_S$ and $\vec{\rho}_I$ the vectors of the population distribution of each type in the metapopulation, system \eqref{edo} can be written as
%\begin{widetext}
\begin{equation}
\label{compact}
\left\{
\begin{array}{l}\displaystyle
 \vec{\rho}\,'_{S} =  \textrm{diag}\big(\tilde{\mu} -
\beta \,c(\rho_k) \frac{\rho_{S,k}}{\rho_k}\big)\, \vec{\rho}_{I} \smallskip\\
\qquad \qquad + (\mathcal{C}-Id)\, \textrm{diag}(D_S(\rho_k))\, \vec{\rho}_{S}\, ,\medskip\\
\displaystyle \vec{\rho}\,'_{I} = \textrm{diag}\big(\beta \,c(\rho_k) \frac{\rho_{S,k}}{\rho_k}-\tilde{\mu} \big)\, \vec{\rho}_{I} \smallskip\\
\qquad \qquad  + (\mathcal{C}-Id)\, \textrm{diag}(D_I(\rho_k))\, \vec{\rho}_{I}\, ,
\end{array}
\right.
\end{equation}
%\end{widetext}
where $\tilde{\mu} = \mu + \delta$, $\mathcal{C}$ denotes the connectivity matrix with $\mathcal{C}_{kk'} = k P(k'|k) / k' $, and $\textrm{diag}(\cdot)$ denotes  a diagonal matrix. $\mathcal{C}$ is non-negative and assumed to be irreducible. For the case of uncorrelated networks, $P(k'|k) = k'p(k')/\langle k \rangle$ and, hence, $\mathcal{C}_{kk'} = k p(k') / \langle k \rangle$, where the brackets stand for the mean value.

\section{Diffusion without epidemics} \label{sec:migration}

In the absence of epidemic ($\vec{\rho}_I = \vec{0}$), the system is driven by the diffusion process. To study the impact of diffusion rates on the population distribution $\rho_k$, we consider the equation
$
\vec{\rho}\,'_{S} = (\mathcal{C}-Id)\, \textrm{diag}(D_S(\rho_k))\, \vec{\rho}_{S}
$
and study the disease-free equilibrium $\vec{\rho}_S^{\, *}$. As $\mathcal{C}$ is irreducible, $v_k=k$, $k=1, \dots , k_{\max}$, is the only positive eigenvector of $\mathcal{C} - Id$ associated to the dominant eigenvalue $\lambda =0$. So, $\rho_k^{\, *}$ satisfies
$D_S(\rho_k^{\, *}) \rho_k^{\, *} = M \, k$,
with $M$ being a suitable constant. Assuming that the total outflow of individuals per patch $F(\rho) := D_S(\rho) \rho$ is continuous, strictly increasing, and satisfies the natural hypothesis $F(0)=0$, the existence and uniqueness of a disease-free equilibrium
\begin{equation}
\label{rho*}
\rho^*_k = F^{-1}(M \,k)
\end{equation}
is guaranteed and, hence, $M$ is computed from the normalizing condition $\sum_k p(k) \rho^*_{k} = \rho^0$. It is worth noting that $\rho^*_k$ does not depend on the conditional probability $P(k'|k)$, so it is independent of the degree-degree correlations, and, moreover, that it increases with $k$ since the total outflow per patch $F(\rho)$ is an increasing function.

To deal with both positive and negative density dependencies, we will assume the following typical nonlinear form for the diffusion rate $D_S$ (see \cite{Mendez} and the references therein)
\begin{equation}\label{diffusion}
D_S(\rho) = D_S^0\, \left(\frac{\rho}{\rho^0}\right)^{\alpha}, \, \alpha > -1 \, ,
\end{equation}
with $D_S^0 >0$ (units of time$^{-1}$), which guarantees the previous hypotheses on the total outflow per patch $F(\rho)$. Regarding to the dimensionless parameter $\alpha$, for $-1~<~\alpha~<~0$ we have the scenario in which the propensity to emigrate is higher in lightly populated patches ($\rho < \rho^0$). On the other hand, an exponent $\alpha > 0$ models the scenario in which the propensity to emigrate is higher in heavily populated patches ($\rho > \rho^0$), and it has been used for modelling animal dispersal in \citep{Murray} (Section 11.3). Finally, $\alpha = 0$ recovers the constant diffusion rate considered in \cite{Saldana2008, Juher2009}.

Taking \eqref{diffusion}, the expression \eqref{rho*} for the disease-free equilibrium reads
\begin{equation} \label{rho*ex}
\rho^*_k= \frac{k^{1/(1+\alpha)}}{\langle k^{1/(1+\alpha)} \rangle} \rho^0.
\end{equation}
Note that this power-law distribution arises independently of the network topology. Moreover, other types of profiles are possible by just changing the form in \eqref{diffusion}. It is worth mentioning that, in terms of our notation, the same profile as \eqref{rho*ex} is found in \citep{Colizza2008} by assuming a constant overall diffusion rate $D^0$ from a patch, but diffusion rates between pairs of patches depending on the degrees of the involved nodes. Therefore, more than one mechanism of dispersal can explain power-law profiles of $\rho^*_k$ (see Discussion).

\begin{figure}[t]
\begin{center}
\includegraphics[scale=0.45]{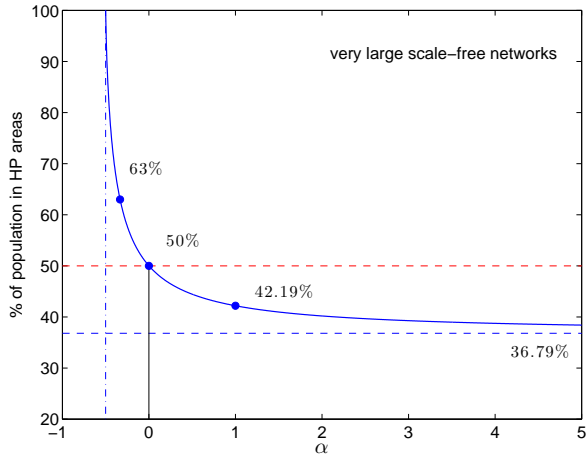}
\caption{(Color online) Percentage of population in heavily populated (HP) patches, $\left( (\omega - 1)/\omega \right)^{\omega-1}\times 100\%$, with $\omega= (\gamma-1)(1+\alpha)> 1$, as a function of the dimensionless migration exponent $\alpha$ in \eqref{diffusion}, and degree distribution $p(k) \sim k^{-\gamma}$ with $k_{\min} = 3$ and $\gamma= 3$. Solid dots correspond to the profiles for $\alpha = -1/3$  (heavily populated scenario), $\alpha = 0$ when the ratio of the individuals living in heavily/lightly populated patches is one, and $\alpha = 1$ (lightly populated scenario).
}\label{fig:profile-urban}
\end{center}
\end{figure}

%\subsection{HP vs non-HP patches} \label{sec:rural-urban}

Beyond the richer variety of population profiles found so far, it is interesting to know to what extent these stationary profiles correspond to metapopulations whose individuals live mostly in heavily populated patches or, on the contrary, in lightly populated ones. Precisely, we say that a patch of connectivity $k$ is heavily populated (HP) if its population is greater than the average number of individuals per patch in the metapopulation, i.e., when $\rho_k > \rho^0$. Otherwise, a patch is said to be lightly populated ($\rho_k \leq \rho^0$).

When the disease-free equilibrium is computed from \eqref{rho*}, HP patches turn out to be those with connectivity larger than $k^*:= D_S(\rho^0) \rho^0/M$, with $M$ the normalizing constant for the profile.
To illustrate heavily/lightly populated scenarios, we assume the form \eqref{diffusion} for the diffusion rate $D_S$ and a scale-free network with degree distribution $p(k)= \frac{\gamma-1}{k_{\min}}(k/k_{\min})^{-\gamma}$, $k \geq k_{\min}$, $\gamma >2$. From \eqref{rho*ex}, HP patches are those with connectivity $$k > k^* = \left( \frac{\omega}{\omega-1} \right)^{1+\alpha} k_{\min}$$ where $\omega= (\gamma-1)(1+\alpha)$, and the total population in HP patches, in the limit of very large networks, is
\begin{equation}\label{urban}
 \displaystyle N \int_{k^*}^\infty p(k) \rho^*_k\, dk= N \rho^0 \left( \frac{\omega-1}{\omega} \right)^{\omega-1}\; ,
\end{equation}
with  $\omega > 1$ to assure the convergence of the integrals. In the situation above and for $\omega=2$, the percentages of individuals living in both types of patches are equal to $50\%$. That would correspond, e.g., to the case of constant diffusion rate $\alpha = 0$ and $\gamma=3$ as in \cite{Juher2009}. However, other migration patterns allow us to describe a wider spectrum of population distributions, specifically, those with a percentage of population living in HP patches higher ($\omega<2$) or lower ($\omega>2$) than $50\%$, see Fig.~\ref{fig:profile-urban} for the case $\gamma=3$ and varying $\alpha$. For any exponent $\gamma>2$ we get analogous pictures as in Fig.~\ref{fig:profile-urban}. In particular, from \eqref{urban} and for a fixed exponent $\gamma>2$ of the degree distribution, the migration exponent $\alpha$ can be used as a tuning parameter to shape the profile to a metapopulation with a given percentage of population in HP patches.

However, the percentage of population living in HP patches, namely, $\left( (\omega - 1)/\omega \right)^{\omega-1} \times 100\%$,  always lies in $(36.79\%, 100\%)$. This is due to the fact that this percentage is decreasing in  $\omega$, with the limit values $100\,e^{-1} \%$ and $100\%$ obtained when $\omega \to \infty$ and $\omega\to 1$, respectively. Notice that it is possible to obtain $100 \%$ of the total population in highly populated patches because the computation in \eqref{urban} assumes infinite network sizes.

\begin{figure*}
\begin{center}
\begin{tabular}{lr}
\includegraphics[scale=0.45]{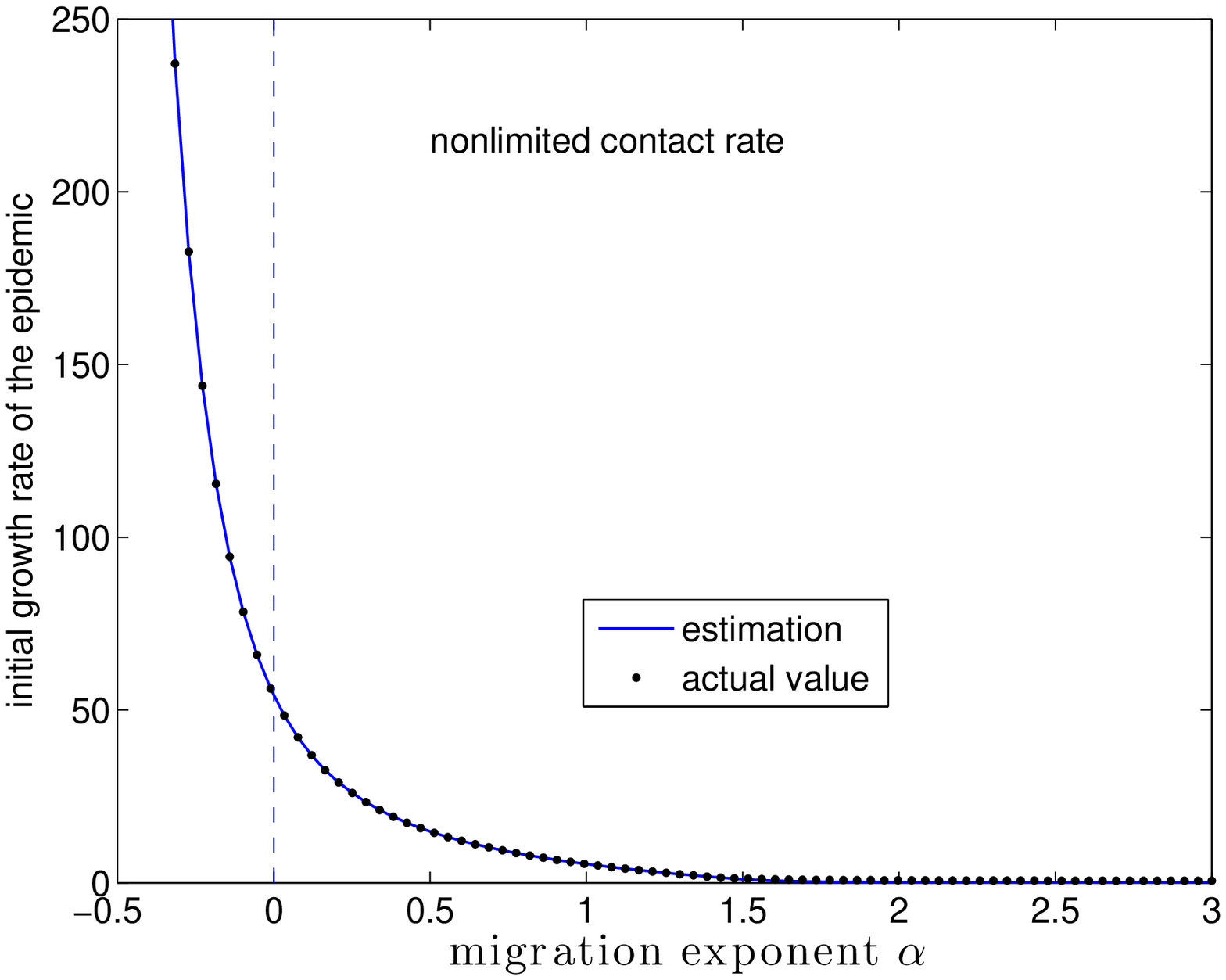} & \includegraphics[scale=0.45]{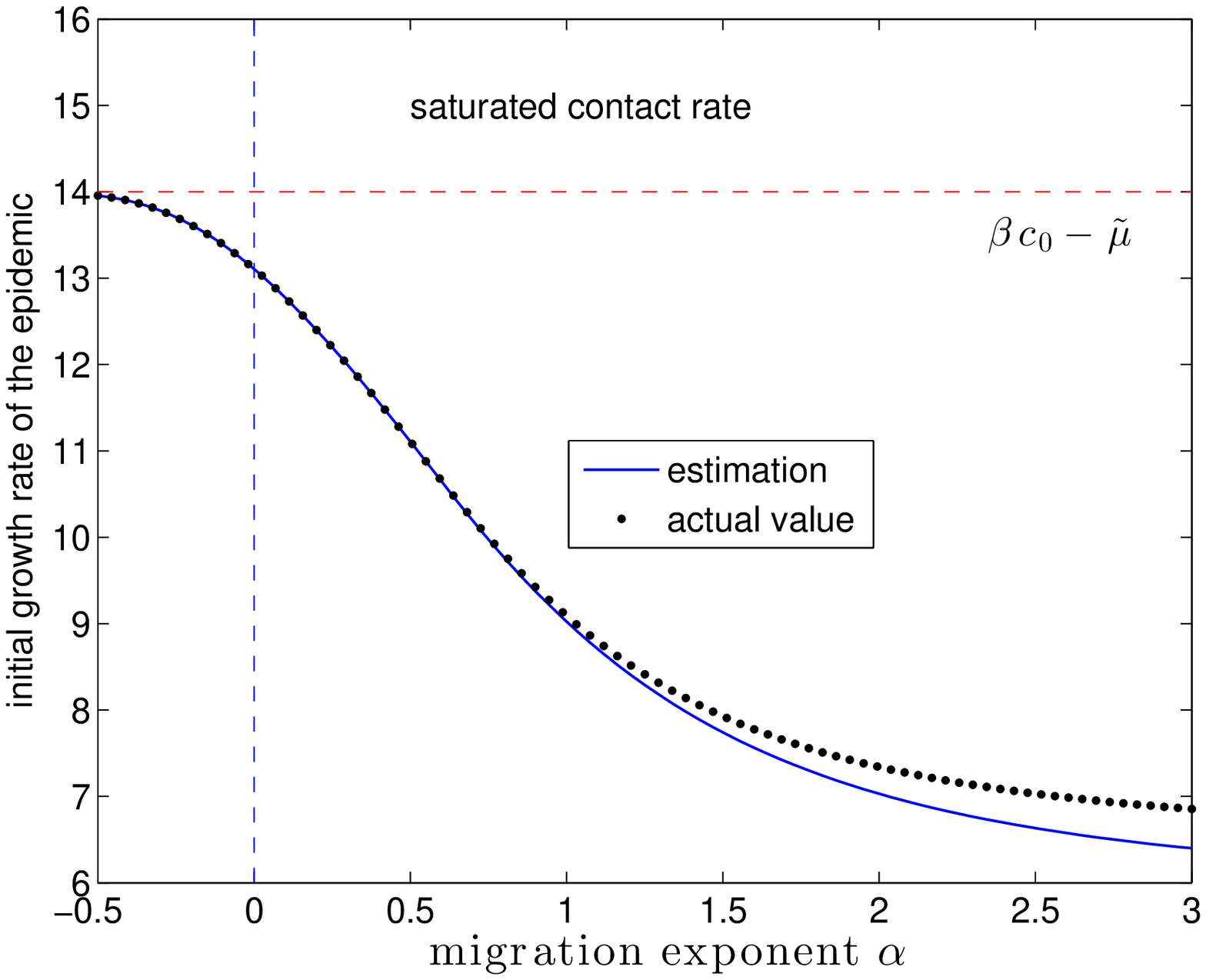}
\end{tabular}
\caption{(Color online) Comparison of the initial epidemic growth rate (time$^{-1}$) for two contact patterns: $c(\rho) = 100 \rho / \rho^0$ (left) and $c(\rho) = 1000 \rho /(\rho^0+\rho)$ (right). Growth rate computed as its actual value (dots) given by the dominant eigenvalue $\lambda_1$ of the Jacobian matrix ($k_{\min}=3, k_{\max}=220$), and its estimation (solid line) given by left-hand side of  \eqref{growth}. Results obtained for uncorrelated scale-free networks ($\gamma = 3, N=5000$ nodes) and parameters:  $\beta = 0.015$ (dimensionless); $\tilde{\mu}=1$, $D_S(\rho)= D_I(\rho)= 0.5 (\rho/ \rho^0)^\alpha$ (units of time$^{-1}$), and $\rho^0=100$ individuals.} \label{fig:initial}
\end{center}
\end{figure*}

\section{Early stage of the epidemic} \label{sec:epidemics}

Now, let us focus on the effect of the migration patterns on the onset of the epidemic. The initial epidemic growth rate is given by the dominant eigenvalue (i.e. spectral bound) $\lambda_1$ of the Jacobian matrix of \eqref{compact} at the disease-free equilibrium. This matrix can be written in blocks as:
\begin{widetext}
\begin{equation}\label{Jacobian}
    \left(
    \begin{array}{cc}
     (\mathcal{C}-Id) \cdot \textrm{diag}\left( F'(\rho_k^*) \right) &  - \textrm{diag}\left( \beta c(\rho^*_k) - \tilde{\mu} \right) \\
     0 & (\mathcal{C}-Id) \cdot \textrm{diag}\left( D_I(\rho_k^*) \right) + \textrm{diag}\left( \beta c(\rho^*_k) - \tilde{\mu} \right)
    \end{array}
  \right),
\end{equation}
\end{widetext}
where the total outflow per patch $F(\rho)=D_S(\rho) \rho$ is assumed to be differentiable. An exponential initial growth of the infectious population occurs when the disease-free equilibrium becomes unstable, i.e., when  $\lambda_1 > 0$. The block structure of the Jacobian matrix and the fact that the dominant eigenvalue of the first block is $0$, imply that $\lambda_1$  is given by the dominant eigenvalue of the fourth block when the latter is positive.

For a constant contact rate, $c(\rho)=c_0$, the dominant eigenvalue is equal to $\lambda_1= \max \{0, \beta c_0 - \tilde{\mu} \}$ and it turns out to be independent of the migration pattern. On the contrary, when the contact rate is density dependent, a sufficient condition for $\lambda_1 > 0$ to be fulfilled is that
\begin{equation}
\max_k \{\beta c(\rho_k^*) - \tilde{\mu} - (1-P(k|k)) D_I(\rho_k^*)\} > 0
\label{growth}
\end{equation}
which follows after grouping all diagonal terms in the fourth block (see appendix A.8 in \cite{Thieme} for details). So, left-hand side of \eqref{growth} is a lower bound of $\lambda_1$ since it neglects off-diagonal terms of the connectivity matrix.

To have a global description of the role of the migration, we have computed the initial growth of the epidemic $\lambda_1$ as the largest eigenvalue of the fourth block of the Jacobian matrix, for different migration exponents $\alpha$ and we have checked the goodness of its estimation given by the left-hand side of \eqref{growth}. Fig.~\ref{fig:initial} shows the fit for two contact patterns, namely, nonlimited and saturated number of contacts per unit of time. The first is the standard mass action $c(\rho)=c_0 \rho$ and in the second we consider a saturated contact rate of the form $c(\rho) = c_0 \rho /(\rho^0+\rho)$, $c_0>0$, which comes from the assumption that the time an individual is available for contacts is limited \citep{Thieme}. In both cases shown in Fig.~\ref{fig:initial}, the initial growth rate of the epidemic decreases as the migration exponent increases.

Recalling the definition of the basic reproduction number of a local population as the number of secondary infections produced by a typical individual in a wholly susceptible population, and computed as \textit{infection probability} $\times$ \textit{contact rate} $\times$ \textit{average infectious period}, we can rewrite \eqref{growth} as
\begin{equation} \label{dominant}
\max_k \frac{\beta c(\rho_k^*)}{\tilde{\mu} + (1-P(k|k)) D_I(\rho_k^*)} > 1 \, .
\end{equation}
So, for a given $k$, the ratio in \eqref{dominant} is a lower bound of the basic reproduction number of the populations living in patches with connectivity $k$, since the immigration from patches with connectivities $k' \ne k$ is neglected. Note that the factor $1-P(k|k)=\sum_{k' \neq k} P(k'|k)$ in the denominator accounts for those individuals who emigrate to patches with other connectivities. Therefore, local epidemic outbreaks will undoubtedly take place in those patches with connectivity $k$ where the ratio in \eqref{dominant} is larger than one.

\begin{figure*}
\begin{center}
\begin{tabular}{lr}
\includegraphics[scale=0.45]{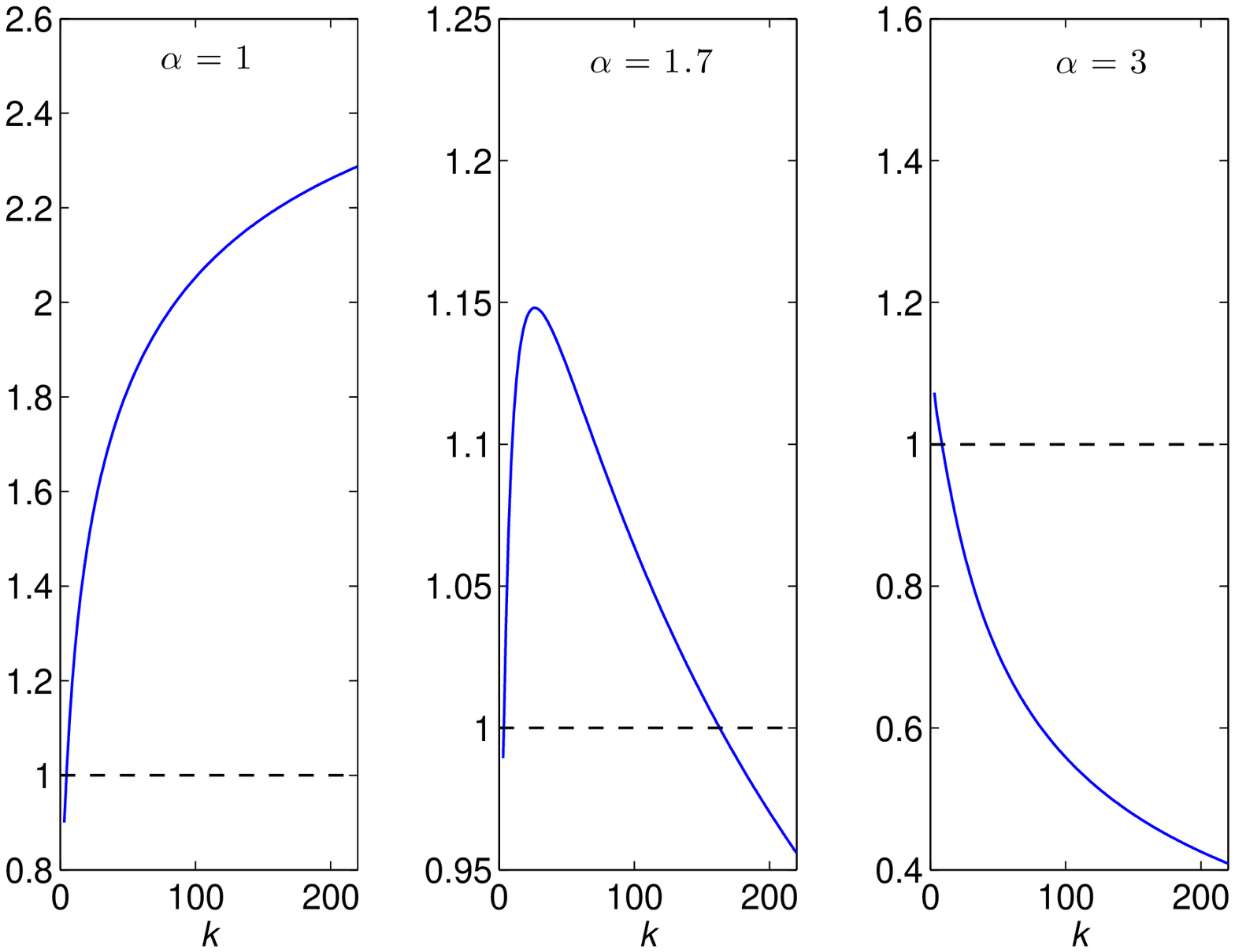} & \includegraphics[scale=0.45]{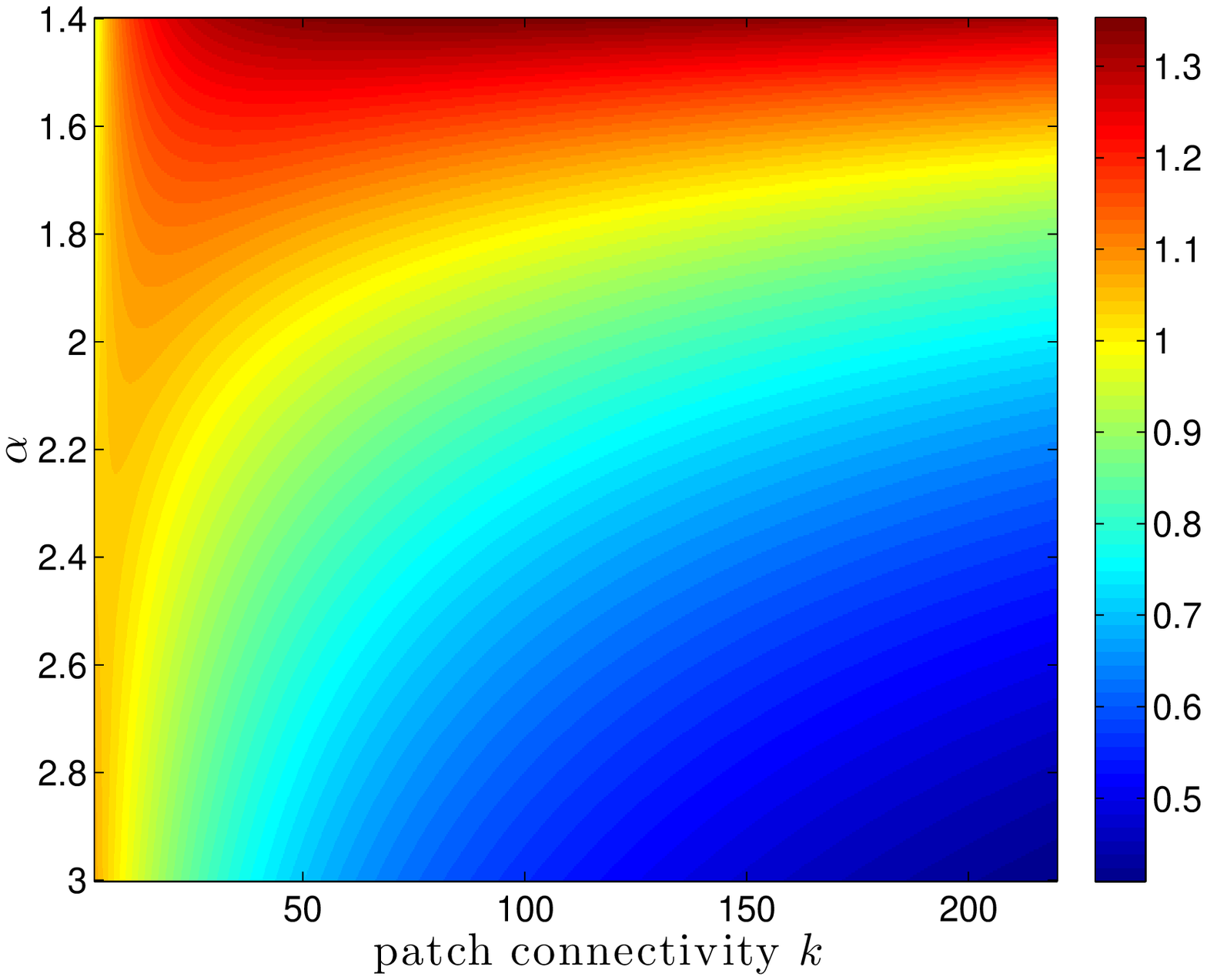}
\end{tabular}
\caption{(Color online) Dimensionless ratio in \eqref{dominant} for uncorrelated scale-free networks ($\gamma = 3$ and $k_{\min}=3$) and diffusion $D_S(\rho)= D_I(\rho)= D^0 \cdot (\rho/ \rho^0)^\alpha$. Left panel, from left to right: migration exponent $\alpha=1$, 1.7, and 3. Right panel: filled contour plot of the ratio in \eqref{dominant} showing that the maximum value decreases as $\alpha$ increases for the considered range of the parameter values. Parameters: $\beta = 0.015$ (dimensionless); $\tilde{\mu}=1$, $c(\rho)=100 \rho / \rho^0$, $D^0= 0.5$ (units of time$^{-1}$), and $\rho^0=100$ individuals.}\label{fig:cota}
\end{center}
\end{figure*}

In contrast to the case of constant diffusion \cite{Juher2009}, for density-dependent contact rates $c(\rho)$ and for $D_I(\rho) = \tau D_S(\rho)$, $\tau>0$, with $D_S$ of the form \eqref{diffusion}, numerical computations show that the maximum in \eqref{dominant} is not necessarily attained at populations with the largest connectivity. In some cases, this maximum is attained at the minimum degree $k_{\min}$, and, in others, at an intermediate value of $k$ (see left panel in Fig.~\ref{fig:cota} for an illustration of these cases). In addition, the right panel of Fig.~\ref{fig:cota} is a plot of the ratio in \eqref{dominant} as a function of both the degree $k$ and the migration exponent $\alpha$, and shows that the higher the migration exponent $\alpha$, the lower the maximum value \eqref{dominant}, for the considered range of parameters values.

\section{Discussion}

Using the modelling framework introduced in \cite{Saldana2008}, a density-dependent diffusion rate $D(\rho)$ has been considered to model positive and negative density dependencies of individuals' migratory movements within a metapopulation \cite{Matthysen}. In order to analyse the effect of these dependencies on the dispersal process, $D(\rho)$ has been assumed to be a power-law function of $\rho$ with an exponent $\alpha > -1$.  This lower bound for the allowed values of $\alpha$ guarantees both that the flow of individuals leaving a patch, $D(\rho) \rho$, strictly increases with its population $\rho$, and that $D(\rho) \rho \to 0$ as $\rho \to 0$ (nobody migrates from where nobody lives).

Under these two natural hypotheses on $D(\rho) \rho$, the disease-free equilibrium is determined by the diffusion process (which does not need to be necessarily described in terms of a power-law diffusion rate) and it is computed from the eigenvector associated to the largest eigenvalue of the connectivity matrix. With this respect, it is interesting to observe that if these hypotheses are violated as by assuming that the number of individuals traveling out of a patch, $D(\rho) \rho$, is independent from its population size $\rho$, the diffusion process does not determine the profile of the disease-free equilibrium. Such an assumption, indeed, leads to a diffusion rate $D$ inversely proportional to the patch population $\rho$, and to an equilibrium which is uniquely determined by the initial distribution $\rho_k(0)$ (see Sect. 3.2 in \citep{Colizza2008}).

On the other hand, when $D(\rho) \sim \rho^\alpha$ ($\alpha > -1$), the disease-free equilibrium is $\rho^*_k \sim k^{1/(1+\alpha)}$ independently of the network topology. So, a power-law profile for $\rho^*_k$ has been obtained under the only assumption of nonlinear diffusion. Remarkably, the same type of profile has been obtained for uncorrelated networks in \citep{Colizza2008} under a different physical principle, namely, assuming that the migratory flow $\omega_{kk'}$ between an origin patch of degree $k$ and a destination patch of degree $k'$ is $w_{kk'} = w_0(k k')^\theta$, $\theta > 0$. From these flows, the diffusion rate between nodes of degree $k$ and $k'$ is taken to be equal to $D^0 w_{kk'}/T_k$ with $T_k = k\sum_{k'} P(k'|k) w_{kk'}$, which ensures a \textit{constant} diffusion rate $D^0$ from any patch in the metapopulation \citep{Colizza2008}.

Which of these physical assumptions is mainly at work in a \textit{random} dispersal process depends on the species we are interested in. In general, it is expectable that many insects and other animal species experience only local environmental conditions and, hence, density dependencies at the origin location would be the main factor in random dispersal within a metapopulation. On the contrary, humans migrate with knowledge of the political, economic and environmental conditions in the potential arrival patches. Under this circumstance, it is suitable to assume that the migratory flow between two patches depends on features of the involved patches, and, in fact, it is what is assumed in many human mobility models \cite{Simini2012, Wilson}. So, distinct physical mechanisms can be behind the observed scaling laws in the topological features of mobility networks \cite{Brockmann2009}.

A way to quantify the effect of different density dependencies is by computing the percentage of individuals living in heavily populated patches (i.e., above the average population per patch). This aggregate measure of the population over the whole metapopulation is a complementary description to that given by $\rho^*_k$. In particular, for infinite scale-free networks and assuming $D(\rho) \sim \rho^\alpha$, this percentage turns out to be bounded from below by $36.79 \%$, which follows as a limit case of the set of power-law distributions, e.g. when the migration exponent $\alpha$ is large enough, i.e., when there is a high propensity to emigrate from heavily populated patches.

%The short-term outbreak situation is given by the linearisation of the system around the disease-free equilibrium and the meaningful expression \eqref{dominant}, or equivalently \eqref{growth}, is an accurate sufficient condition for the epidemic outbreak to happen in the metapopulation. This formula is general in the sense of determining the epidemic spreading for any network topology and degree-degree correlation, any migration pattern, and any contact pattern among individuals. When the contact rate among individuals of a local population is constant, the migration does not play any role. Instead, for density-dependent contact rates, our analysis shows that migration flows play a crucial role in the spread of infectious diseases.

From an epidemiological point of view, the main novelty of our results is to show that migration patterns determine where epidemic outbreaks can take place with higher probability. In particular, for large enough migration exponents, the onset of an epidemic may not be triggered by infectious individuals living in large size populations, as it could be expected from the fact that $\rho^*_k$ increases with the connectivity of the sites. This happens when individuals, also the infectious ones, have a high propensity to move from heavily populated areas. Such a higher number of infectious individuals arriving at patches with lower population sizes makes more likely the occurrence of an outbreak in these patches. But, at the same time, their lower sizes lead to lower values of the basic reproduction number because the contact rate $c(\rho)$ increases with $\rho$ (see Fig. \ref{fig:cota}). Therefore, in contrast to what is usually claimed about the role of the long-range mixing in exacerbating epidemics \cite{EPCH}, we have seen that strengthening the migration from heavily populated areas can help to contain an epidemic by causing its appearance in less populated areas, for which the value of the basic reproduction number is significantly smaller.

%\begin{acknowledgments}
This work is partially supported by grants MINECO MTM2011-27739-C04-03 and MTM2014-52402-C3-3-P (Spain). The authors are members of the Catalan research group 2014 SGR 1083.
%\end{acknowledgments}

%\bibliography{References-PRE-2015}

%

\end{document}